\documentclass[epj]{svjour}
\usepackage{graphics}
\begin{document}
\title{Two-particle rapidity correlations between relativistic particles in central collisions of $^{197}$Au nuclei in emulsion at 11.6 A GeV/c}
\author{U. U. Abdurakhmanov\inst{1} \and K. G. Gulamov\inst{1} \and V. Sh. Navotny\inst{1}
}                     
%
%
\institute{Institute for Physics and Technology, Fizika-Solntse Research and Production Association,
Uzbek Academy of Sciences, ul. G. Mavlyanova 2b, Tashkent, 700084 Republic of Uzbekistan}
\date{Received: date / Revised version: date}
%
\abstract{
It is shown that in central collisions of  $^{197}$Au nuclei with heavy emulsion nuclei at 11.6 $A$GeV/c two-particles pseudorapidity correlations for produced particles in terms of correlation functions demonstate predominantly long-range behaviour in contrast to nucleon-nucleon interactions. The experimental data are compared with calculations based on the FRITIOF-M model and the model of independent emission of particles. 
} 
\maketitle
\section{Introduction}
\label{intro}
Interest to the study of relativistic nucleus-nucleus collisions is caused by many reasons, such as the possibility to study the nuclear fragmentation processes or the processes of propagation of newly born particles ("in statu nascendi") through the nuclear medium. Particular attention is paid to issues related to the collective properties of the particles produced. Since the energy density can reach high values, the question arises on the possibility of an ensemble of produced particles to manifest some collective behaviour (the quark-gluon plasma is the primary goal here), especially in light of the fact that in such collisions the bounce-off effects are clearly observed \cite{ref1}. It can be assumed that the collective properties of an ensemble of produced particles may also demonstrate themselves in two-particle and multiparticle correlations. Recent publications \cite{ref2,ref3,ref4} suggest that rapidity correlations may be considered as a useful tool for distinguishing between different mechanisms of multiparticle production in relativistic heavy ion collisions.

This paper presents experimental data on two-particle correlations among pseudorapidities of relativistic particles produced in central interactions of $^{197}$Au nuclei with emulsion nuclei at 11.6 $A$GeV/c.

\section{Experimental Data}
\label{sec:2}
The experimental data are accumulated in the framework of E-863 experiment of the EMU-01 collaboration  \cite{ref5,ref6}. Emulsions were irradiated by $^{197}$Au nuclei at 11.6 A GeV/c on the AGS accelerator of the Brookhaven National Laboratory (Upton, United States). A beam of gold ions with density $5 \cdot 10^3 \; nuclei/cm^2$ with an admixture of foreign ions of no more than $2\%$ was used for these purposes.

For the analysis the events of inelastic incoherent interactions of gold nuclei in emulsion were collected, the events of electromagnetic nature were excluded from consideration. In accordance with emulsion technique, the secondary charged particles in events were divided into different groups:

black or $b$-particles, mainly consisting of protons from the target nuclei with momenta  $ p \leq  0.2 GeV / c$ and also heavier nuclear fragments;

gray or $g$-particles corresponding to protons with the momenta $0.2 \leq p \leq 1 GeV/c$; they mainly consist of protons - fragments of the target nuclei, contribution of  slow pions does not exceed several percent. Black and gray particles may be combined into a group of strongly ionizing  $h$-particles with a multiplicity $n_h = n_b + n_g$;

shower or $s$-particles - singly charged particles with  speed $\beta \geq 0.7$, mainly consisting  of produced particles and the singly charged fragments of the projectile nucleus;

projectile fragments - fast particles with charge $Z \geq 2$ and  ionization  $I/I_0 \geq 4$, not changing at long distances from the point of interaction in emulsion, where $I_0$  is the ionization of a singly charged relativistic particle.

For all the above types of particles their multiplicity and the emission angles were determined. For the analysis of angular distributions of $s$-particles we use pseudorapidity: 
\begin{eqnarray}
\label{u01}
\eta = - \log \tan \frac{\theta}{2}
\end{eqnarray}
where $\theta$ - is the emission  angle of  the $s$-particle. For pions pseudorapidity is related  with true rapidity  by a simple equation:
\begin{eqnarray}
\label{u02}
\sinh \eta = \frac{m_T}{p_T} \sinh y, \;\;\; \; where \;\; m_T^2 = m^2 + p_T^2.
\end{eqnarray}
Statistics of the experiment  consists of  $1057$ incoherent inelastic interactions of gold nuclei in emulsion. More information about the experiment is given in \cite{ref7}.

As a measure of the centrality of interactions with respect to the target nucleus we use conventional in emulsion technique criterion $n_h \geq 8$, which means that we effectively select collisions of gold nuclei with heavy emulsion nuclei. The number of such events is equal to $450$ from a total number of $1057$ inelastic interactions collected

As a measure of centrality of collisions with respect to the projectile, we use the number $n_\alpha$ of relativistic doubly charged particles - projectile fragments, consisting mostly of $\alpha$-particles. All central events considered by us were divided into three groups according to the number of doubly charged particles in them. Table 1 presents the average multiplicity of $s$-particles in these groups. One can see that the average multiplicity of shower particles in these groups increases with the decreasing number of doubly charged particles from a projectile nucleus, indicating that the number of doubly charged particles can indeed be considered as a measure of the centrality of interactions considered. From this point of view events with $n_h \geq 8$ and $n_\alpha = 0-2 $ belong to the most central collisions of $^{197}$Au nuclei with heavy emulsion nuclei considered by us.

\begin{table}
\caption{The average multiplicities of s-particles for events with $n_h \geq 8$ and different $n_\alpha$.}
\label{tab:1}       
\begin{tabular}{|c|c|c|c|}
\hline
$n_\alpha$ &  0 - 2 & 3 - 4 & $\geq 5 $ \\
\hline
$\langle n_s \rangle$  & 147.92 $\pm$ 12.76 & 127.94 $\pm$ 7.88 & 113.86 $\pm$ 3.94 \\
\hline
\end{tabular}
\end{table}

\section{Correlation functions and Models}
\label{sec:3}
For study of pseudorapidity correlations between produced particles we use the technique of  inclusive correlation functions. Two-particle correlation function is defined as
\begin{eqnarray}
\label{u03}
C_2(\eta_1, \eta_2) = \rho_2(\eta_1, \eta_2) - \rho_1(\eta_1) \rho_1(\eta_2),
\end{eqnarray}
where  $\rho_2(\eta_1, \eta_2)$, $\rho_1(\eta)$  are the one- and two-particle pseudorapidity distributions, respectively
\begin{eqnarray}
\label{u04}
\rho_1(\eta_1)= \frac{1}{\sigma} \frac{d\sigma}{d\eta} \;, \; \; \; \rho_2(\eta_1, \eta_2)=  \frac{1}{\sigma} \frac{d^2\sigma}{d\eta_1 d\eta_2} \;.
\end{eqnarray}
In our analysis we use the normalized two-particle correlation function 
\begin{eqnarray}
\label{u05}
R_2(\eta_1, \eta_2) = \frac{\rho_2(\eta_1, \eta_2)}{\rho_1(\eta_1) \rho_1(\eta_2)}-1,
\end{eqnarray}
which is less sensitive to the details of one-particle inclusive distributions.

It is well-known (see, e.g.  \cite{ref8,ref9,ref10} and references therein) that the magnitude of correlation functions depends not only on the strength of true dynamic correlations between particles in an event, but also on more general factors, such as the shape of multiplicity distribution of produced particles and/or the dependence of inclusive distributions on the multiplicity, i.e. factors related with inhomogeneous nature of ensembles of inclusive events.  For example, it follows from normalization of inclusive distributions that 
\begin{eqnarray}
\label{u06}
\int\!\!\!\!\int \! C_2(\eta_1, \eta_2)d\eta_1 d\eta_2 =\langle n(n-1) \rangle - \langle n \rangle^2.
\end{eqnarray}
For nucleus-nucleus interactions the inhomogeneous nature of ensembles of events is obvious even for the final states with the same $n_s$ because nuclear interactions with different geometry and therefore different number of intranuclear nucleon-nucleon collisions may lead to the same multiplicity $n_s$ of produced particles. Therefore, when analysing experimental data on correlation functions, it is important to  evaluate contributions of both trivial and truly dynamic effects to their magnitude and shape, which is impossible to do in a model independent way.

The experimental data of the present paper were compared with results of Monte-Carlo simulations in the framework of two models. 

 In the framework of a phenomenological model  of independent emission (IEM) of $s$-particles \cite{ref10,ref11,ref12} we assume that: i) multiplicity ($n_s$) distributions of simulated events in each one of subensembles (with different $n_\alpha$, for example) reproduce the experimental distributions in the real group (subensemble) of events; ii) one-particle pseudorapidity distributions of $s$-particles in each one of simulated subensembles of events (within, for instance, the fixed range of $n_s$ and $n_\alpha$) reproduce the experimental "semiinclusive" distributions for the same $n_s$ and $n_\alpha$ ; iii) emission angles of $s$-particles in each one of simulated events are statistically independent. 

Thus, events simulated in the framework of IEM reproduce experimental multiplicity and angular distributions of $s$-particles for each one of considered group (subensemble) of events but the emission angles of these particles are statistically independent. The model by definition does not introduce any dynamic correlations among $s$-particles. It can be used to  evaluate the contribution of effects related to multiplicity and angular distributions to the correlation functions. Of course, the model does not take into account the energy-momentum conservation which influence on correlation functions was found \cite{ref10} to be negligibly small for high-multiplicity interactions.

In practice we have simulated artificial events  in  which  emission angles of $s$-particles were generated on the basis of the experimental angular distributions for the given multiplicities $n_\alpha$ (and  $n_s$). Then events with different $n_s$ were mixed up with statistical weights corresponding to the experimental multiplicity distribution of $s$-particles. Altogether we have simulated more than $10^5$ events so that the statistical errors are much less than the experimental ones.
\begin{figure*}
\resizebox{0.9\textwidth}{!}{
  \includegraphics{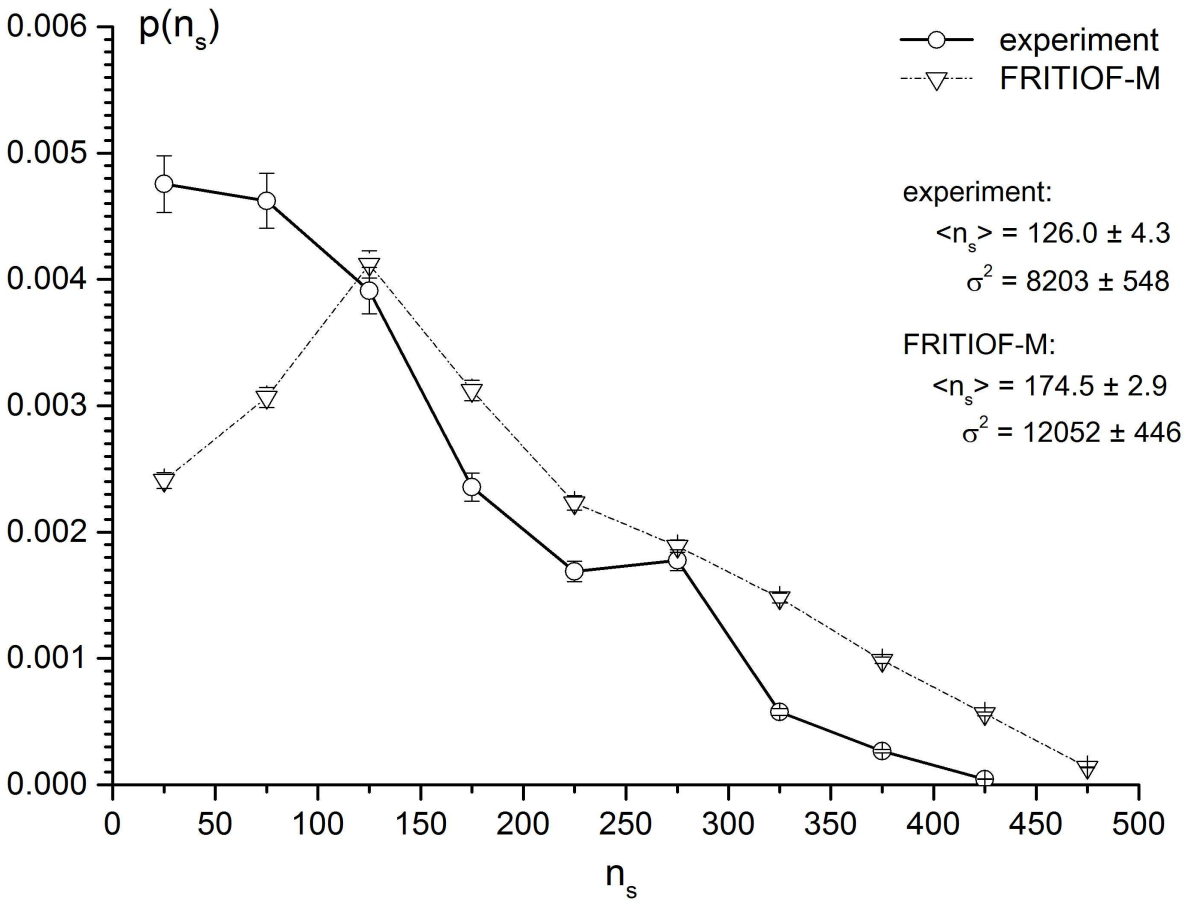}}
\caption{Multiplicity distribution for central events.}
\label{fig:1}       
\end{figure*}

The experimental data were also compared with results of Monte-Carlo calculations performed  in the framework of a modified FRITIOF-M model \cite{ref13}. In this model the first stage of collisions is simulated following the phenomenological approach used in the well-known original FRITIOF \cite{ref14,ref15}, which integrates phenomenological string model for nucleon-nucleon collisions with the Glauber theory for nucleus-nucleus interactions.  The next stage of an interaction related with disintegration of residual nuclei is considered in this approach  following the statistical nuclear multifragmentation model \cite{ref16}. No cascade processes are allowed in the model. Like original FRITIOF, the FRITIOF-M takes into account production and decay of intermediate resonances mostly in the central region of nucleus-nucleus collisions. These resonances can be a source of short-range rapidity correlations  between  $s$-particles of the final state. As regards residual  nuclei,   they undergo  statistically isotropic decays  in their own rest frames, but the model does not introduce any additional dynamic correlations.  More than 5000 events  generated in the framework of this model for interactions of gold nuclei in emulsion at 11.6 A GeV/c  were processed following the same rules and criteria as the experimental ones. It may be noticed that the statistical errors for model calculations are considerably less than for the experimental data.

\section{Experimental results}
\label{sec:4}

In Figure 1 we show multiplicity distribution of $s$-particles in central ($n_h \geq 8$) collisions of gold nuclei with heavy emulsion nuclei at 11.6 A GeV/c in comparison with corresponding results  for the FRITIOF-M model. We see that the experimental $n_s$-distribution is very broad and it can not be reproduced by the FRITIOF-M model \cite{ref13}. In fact, the FRITIOF-M represents by itself an attempt to combine the original FRITIOF, which describes satisfactorily the data on produced particles in hadron-hadron and hadron-nucleus interactions, with multifragmentation model, which describes a number of characteristics of nuclear fragmentation processes. One can conclude that this attempt is not very successful, at least for interactions considered.

In Figure 2 we show the values of the correlation function $R_2(\eta_1, \eta_2)$  at $\eta_1 = \eta_2$, reflecting the magnitude of short-range correlations between $s$-particles, for central collisions of $^{197}$Au nuclei with heavy emulsion nuclei  with different numbers of doubly charged projectile fragments in the final state. We see, first of all, that  $R_2(\eta_1, \eta_2)$  at $\eta_1 = \eta_2$  shows almost no dependence on the pseudorapidities of $s$-particles for all three groups of events and the value of $R_2$ is almost the same in fragmentation and central regions of heavy ions collisions. For the groups of more central interactions  with  $n_\alpha = 0-2$ and $n_\alpha = 3-4 $  the experimental values of  $R_2(\eta_1, \eta_2)$  are approximately equal to zero, which corresponds to the absence of any significant short-range correlations between produced particles in the experiment. For more peripheral with respect to the projectile  collisions with $n_\alpha \geq 5$ the value of $R_2(\eta_1, \eta_2)$ is less than zero, but also does not depend on $\eta$. 

\begin{figure}
\resizebox{0.45\textwidth}{!}{%
  \includegraphics{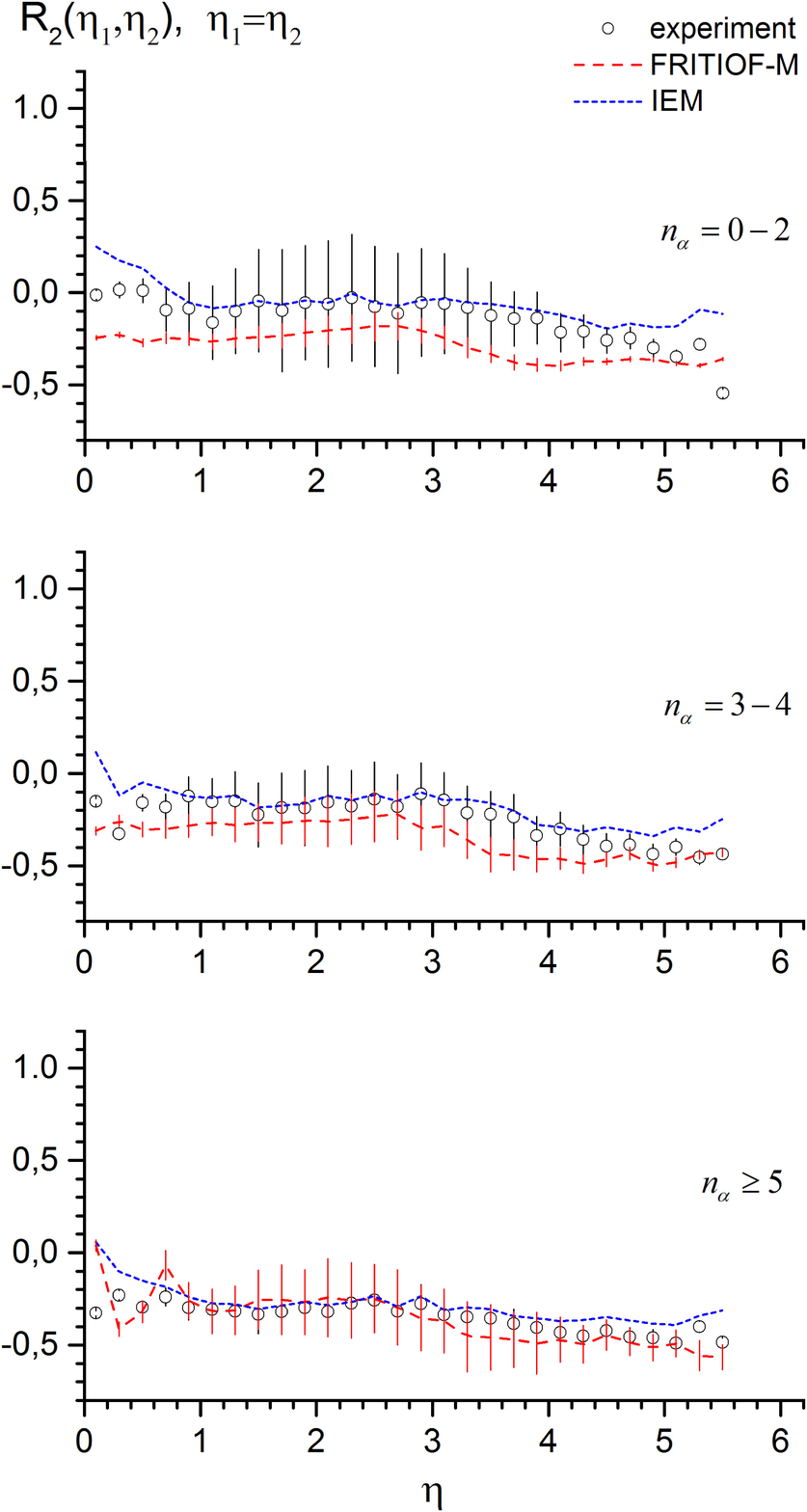}
}
\caption{$R_2(\eta_1, \eta_2)$ for three groups of central events at $\eta_1 =\eta_2$}
\label{fig:2}       
\end{figure}

The general  behaviour  of $R_2(\eta_1, \eta_2)$ in heavy-ions collisions considered is quite different from that in high-energy hadron-nucleon and hadron-nucleus collisions. For example, in high-energy hadron-nucleon interactions correlation functions demonstrate clearly the short-range order, they have a bump in the central region and tend to become zero only in the fragmentation regions \cite{ref9,ref17}. Of course, it is necessary to note that at energies considered by us the central region cannot be separated from fragmentation regions of colliding nuclei. Probably this is one of the reasons why the correlation function $R_2(\eta_1, \eta_2)$  demonstrates flat  $\eta$-dependencies.

Comparison of calculations performed in the framework of FRITIOF-M model with the experimental data shows that the agreement is observed for events with  $n_\alpha \geq 5$, while for central events with $n_\alpha = 0-2$ and $n_\alpha = 3-4 $ the model underestimates the value of $R_2$ at $\eta_1 = \eta_2$.

Model of independent emission of $s$-particles well describes the experimental data for all three groups of events. In other words, dynamic correlations between produced particles do not reveal themselves in experimental data on correlation functions, and deviations of the value of $R_2$ at $\eta_1 = \eta_2$ from zero can be attributed to the influence of the multiplicity and angular distributions of produced particles taken into account in the IEM.

The experimental values of the correlation function $R_2(\eta_1, \eta_2)$  for fixed $\eta_1 = 2.0$ are shown in Figure 3 in comparison with results of calculation in the framework of the FRITIOF-M and IEM.  Here the magnitude of $R_2(\eta_1, \eta_2)$ can be considered as a measure of long-range correlations between $s$-particles for all values of pseudorapidities, except for values of $\eta_1 = \eta_2 = 2.0$. We see that for all of the considered groups of events the experimental values of $R_2(\eta_1, \eta_2)$  demonstrate the presence of significant correlations of the long-range nature. Of course, for $\eta_1 = \eta_2 = 2.0$ the value of $R_2(\eta_1, \eta_2)$  becomes close to zero or even negative in agreement with what we have seen before for short-range correlations in Figure 2. In other words, in this experiment, two-particle correlations between pseudorapidities of produced particles are of the long-range nature, whereas the short-range correlations are suppressed considerably.  

\begin{figure}
\resizebox{0.45\textwidth}{!}{%
  \includegraphics{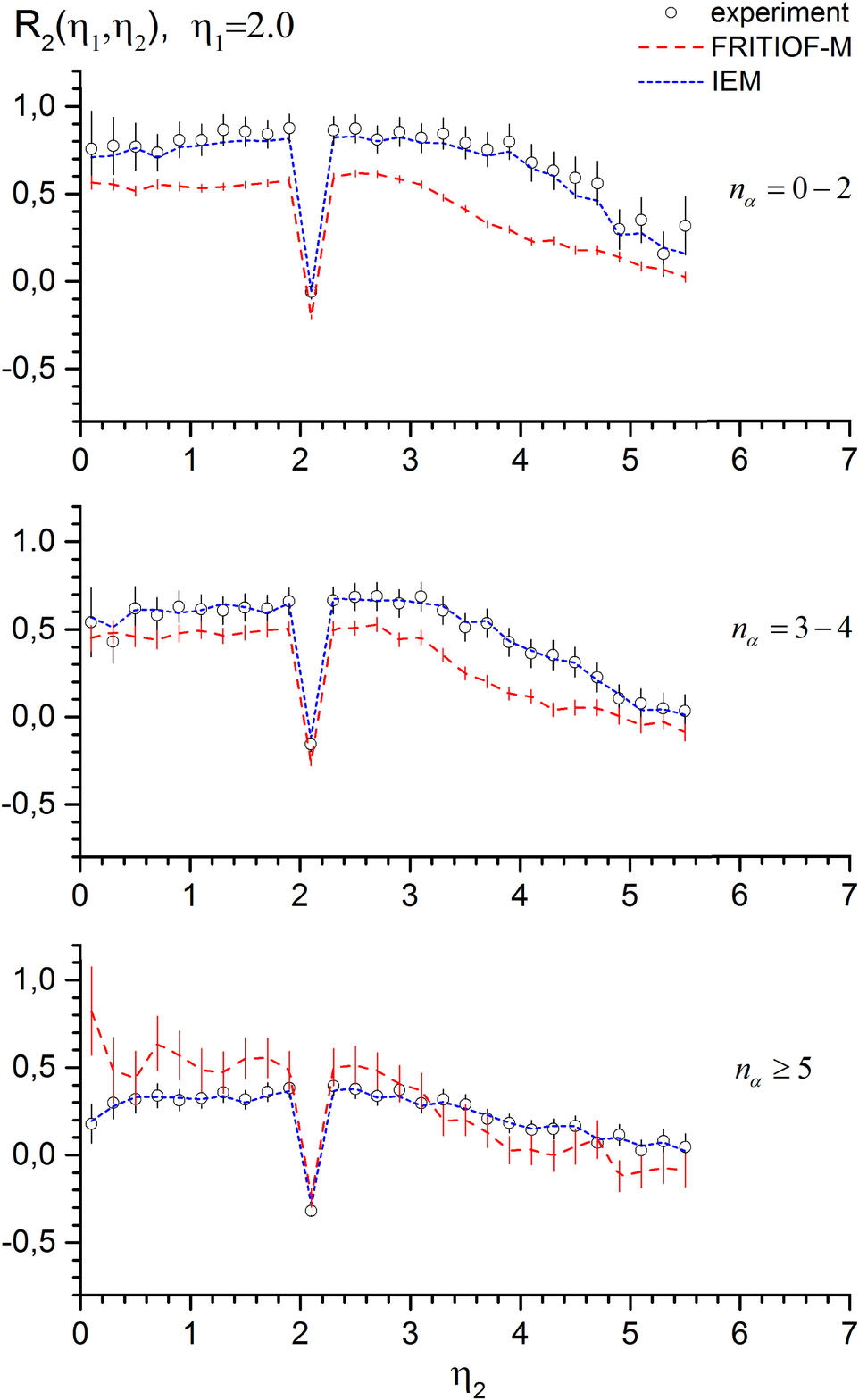}
}
\caption{$R_2(\eta_1, \eta_2)$ for three groups of central events at  $\eta_1 = 2.0$}
\label{fig:3}       
\end{figure}

The values of $R_2(\eta_1, \eta_2)$ for the FRITIOF-M model also show the existence of long-range correlations, although the calculated values are below of the experimental ones for the groups of events with  $n_\alpha = 0-2$ and $n_\alpha = 3-4 $, while for events with $n_\alpha \geq 5$ they generally describe the data.  

Independent emission model well describes the experimental data for all three groups of events. This implies that the long-range correlations observed are not of the dynamic nature but linked to the fluctuations of $s$-particles' multiplicities and angular distributions.

\section{Conclusions}
\label{sec:5}
Analysis of the two-particle pseudorapidity correlations for relativistic particles shows that short-range correlations are strongly suppressed in central collisions of $^{197}$Au nuclei with heavy emulsion nuclei in comparison with high-energy hadron-nucleon interactions. In fact the experimental data do not reveal any noticeable short-range correlations between pseudorapidities of $s$-particles. At the same time the data on the normalized correlation functions demonstrate the presence of considerable long-range correlations. The enhanced contribution of long-range correlations between pseudorapidities of particles produced in hadron-nucleus interactions in comparison with hadron-nucleon collisions observed  earlier \cite{ref11} was linked to an additional source of multiplicity fluctuations associated with multiple collisions of a projectile inside the target nucleus. The same reasoning can be applied to the case of relativistic heavy-ions interactions.

The comparison of experimental data with calculations based on the independent emission model shows that the source of these correlations is not of a dynamic origin, and most probably is related with fluctuations in the number of intranuclear nucleon-nucleon collisions which influences the multiplicity of produced particles. This conclusion is confirmed by the fact that the long-range correlations also appear in the FRITIOF-M model. Taking into account the nuclear geometry FRITIOF-M introduces heterogeneity of events which is in turn reflected in multiplicity and pseudorapidity distributions of final state particles. Numerical deviations of the FRITIOF-M model from the experiment may be the result of the fact that the model does not well describe  multiplicity distributions of produced particles for the central collisions of  gold nuclei with heavy emulsion nuclei.

We are thankful to all members of the EMU-01 collaboration with whom the experimental data were collected and analyzed.

%
%

\end{document}